\documentstyle[preprint,aps]{revtex}
\begin{document}
\draft
\title{
Mapping of the Sutherland Hamiltonian to anyons on a ring}
\author{Shuxi Li and R.K. Bhaduri}
\address{
Department of Physics and Astronomy,
McMaster University,\\
Hamilton, Ontario, Canada L8S 4M1
}
\maketitle

\begin{abstract}
It is demonstrated that the Sutherland Hamiltonian is
equivalent to particles interacting with the vector
statistical interaction
on the rim of a circle, to
within a nonsingular gauge transformation.
\end{abstract}
\vspace {7 mm}
\pagebreak

\narrowtext

Sutherland \cite{Suth71a} obtained a periodic potential by
allowing a pair of particles on the circumference of a
circle to interact any number  of times by an inverse
square potential. Consider $N$ such particles, each of
mass $M$, described by the Sutherland Hamiltonian $H_s$:
\begin{equation}
({2M\over \hbar^2})H_s=-\sum_i^N{\partial^2\over \partial
x_i^2}+\beta({\beta\over 2}-1){\pi^2\over L^2}\sum_{i>j}
\left(\sin {\pi(x_i-x_j)\over L}\right)^{-2}\;.
\label{eq:1}
\end{equation}
In the above, $x_i$ is the coordinate of the $i$th particle,
measured along the arc length of a circle of circumference $L$,
and $\beta$ is a dimensionless constant. The many-body
Hamiltonian $H_s$ is exactly solvable in astonishing detail
\cite{Suth72}. Besides being the direct descendent of the
inverse square potential model of Calogero \cite{Calo69},
$H_s$ has many intriguing connections with other systems
\cite{Simo93,Simo94,Shas93}. Sutherland had himself noted
\cite{Suth71a,Suth71b}, for example, that the square of the
ground state spatial wave function was identical in form
to the joint probability distribution of the eigenvalues
of random matrices. Recently, Simons {\it et. al}
\cite{Simo93,Simo94} have further demonstrated that the
time-dependent density-density correlations
of $H_s$ are of the universal
form that is common in the spectral correlator of disordered
and chaotic quantum systems under arbitrary perturbations.
Thus there seems to be a deep-rooted connection between
integrable and nonintegrable systems through the inverse
square interaction. In this letter, we demonstrate that
the Sutherland Hamiltonian $H_s$ is also equivalent (to
within a gauge transformation), to $N$ particles on a circle
interacting with the so-called ``statistical interaction''
\cite{Lein77,Wilc82,Murt92}. By   constraining the
particles on the rim of a
circle, the two-dimensional anyon problem reduces to a
one-dimensional one, and is integrable.
 In view of the fundamental
role played by $H_s$ (and therefore the integrable anyon
problem), it is natural then to ask about the possible
connection between two-dimensional nonintegrable anyon models
and chaos.

Consider $N$ nonrelativistic particles on a ring of unit
radius in the $x$-$y$ plane, interacting pair-wise through
a statistical type vector interaction. The particles have
charge $e=-|e|$, and mass $M$, and the Hamiltonian is given by
\begin{equation}
H={1\over 2M}\sum_i^N\left({\bf p}_i+
{|e|\over c}{\bf A}_i\right)^2 \;
,\label{eq:2}
\end{equation}
where
\begin{equation}
 {\bf A}_i({\bf r}_i)=\alpha
{{\hbar c}\over |e|}
\sum_{j}^{\prime} {{\hat{\bf z}}\times({\bf r}_i
-{\bf r}_j)
\over{\bigl|{\bf r}_i-{\bf r}_j\bigr|^2}}\;
.\label{eq:3}
\end{equation}
Note that the strength $\alpha$ of the interaction is
dimensionless. For simplicity,
from now on, we put $\hbar=c=|e|=2M=1$.
In Eq.~(\ref{eq:3}), ${\bf r}_i$, ${\bf r}_j$ are radial
position vectors in the $x$-$y$ plane for the $i$th
 and $j$th particles, and
 ${\hat{\bf z}}$ is a unit vector along the $z$-direction.
The prime on the sum in Eq.~(\ref{eq:2})
implies that $j{\ne}i$. Let the vectors ${\bf r}_i$,
${\bf r}_j$ make angles $\phi_i$, $\phi_j$ with the
$x-$axis, and define $\phi_{ij}=(\phi_i-\phi_j)$,
${\bf r}_{ij}=({\bf r}_i-{\bf r}_j)$.
It is then easy to show that,
\begin{equation}
{\bf r}_{ij}
=\;2 \sin{\phi_{ij}\over 2}
({\hat{\bf r}}_i \sin{\phi_{ij}\over 2} +{\hat{\bf e}}_{\phi i}
\cos{\phi_{ij}\over 2} )\;
.\label{eq:4}
\end{equation}
Here ${\hat{\bf r}}_i$, ${\hat{\bf e}_{\phi i}}$ are unit
vectors in the radial and tangential directions at the site
of the $i$th particle. Substituting the expression ~(\ref{eq:4})
for ${\bf r}_{ij}$ in Eq.~(\ref{eq:3}), and simplifying,
one obtains
\begin{equation}
{\bf A}_i={1\over 2}\alpha\sum_j^{\prime}
({\hat{\bf e}}_{\phi i}-
{\hat{\bf r}}_i \cot{\phi_{ij}\over 2})\;
.\label{eq:5}
\end{equation}
Using this in Eq.~(\ref{eq:2}), the Hamiltonian $H$ reduces
to the form
\begin{eqnarray}
H=\sum_i(-i{\partial\over \partial\phi_i}+{\alpha\over 2}(N-1))^2
+{\alpha^2\over 4}\sum_i\sum_j^{\prime}(\sin{\phi_{ij}\over 2})^{-2}
\nonumber\\
+{\alpha^2\over 4}\sum_i\sum_j^{\prime}\sum_k^{\prime\prime}
\cot{\phi_{ij}\over 2}\cot{\phi_{ik}\over 2}-{\alpha^2\over 4}
N(N-1)\;
.\label{eq:6}
\end{eqnarray}
In the above, the sum $\sum_k^{\prime\prime}$ indicates that
 $k{\ne} i$ and $k{\ne} j$. Further simplification
results by noting that the triple sum obeys the identity
\begin{equation}
\sum_i\sum_j^{\prime}\sum_k^{\prime\prime}
\cot{\phi_{ij}\over 2}\cot{\phi_{ik}\over 2}=-{1\over 3}N(N-1)(N-2)\;
,\label{eq:7}
\end{equation}
yielding
\begin{equation}
H=\sum_i(-i{\partial\over\partial \phi_i}+{\alpha\over 2}(N-1))^2
+{\alpha^2\over 4}\sum_i\sum_j^{\prime}(\sin{\phi_{ij}\over 2})^{-2}
-{\alpha^2\over 12}N(N^2-1)\;
.\label{eq:8}
\end{equation}
Note that this ``anyon'' Hamiltonian $H$ does not respect
time-reversal symmetry. It can, however, be transformed to
the Sutherland Hamiltonian $H_s$ (Eq.~(\ref{eq:1})) by
applying a {\it nonsingular} gauge transformation. Let
the eigenequation of $H$ defined by Eq.~(\ref{eq:8})
  be written as $H\Psi=E\Psi$,  applying the gauge
transformation
\begin{equation}
\Psi=\Psi_s\exp (-i{\alpha\over 2}(N-1)\sum_i^N\phi_i)\;.
\label{eq:9}
\end{equation}
Then one obtains
\begin{equation}
{H^{\prime}}\Psi=E\Psi_s\;,
\nonumber
\end{equation}
\begin{equation}
H^{\prime}=-\sum_i{\partial^2\over\partial \phi_i^2}
+{\alpha^2\over 4}\sum_i\sum_j^{\prime}(\sin{\phi_{ij}\over 2})^{-2}
-{\alpha^2\over 12}N(N^2-1)\;
.\label{eq:10}
\end{equation}
Note that $H^{\prime}$ is identical to the Sutherland Hamiltonian
$H_s$ if the constant term
 ${\alpha^2\over 12}N(N^2-1)$ is added to it, and the parameter
$\alpha$ above is related to the coupling strength in
Eq.~(\ref{eq:1}) by the relation
\begin{equation}
\alpha^2={\beta\over 2}({\beta\over 2}-1)\;.
\label{eq:11}
\end{equation}
This follows since $(x_i-x_j)={L\over 2\pi}(\phi_i-\phi_j)$,
and for unit radius $L=2\pi$. Since the entire eigenspectrum of
the Sutherland Hamiltonian $H_s$ is known \cite{Suth72},
the corresponding one for $H$ of Eq.~(\ref{eq:8}) may be
deduced immediately. For example, the ground state energy of
$H$ is given by
\begin{equation}
E_0={\beta\over 24}N(N^2-1)\;,
\label{eq:12}
\end{equation}
with the eigenstate
\begin{equation}
\Psi_0={\sqrt C} e^{-i{\alpha\over 2}(N-1)\sum_i\phi_i}
\prod_{i>j}|e^{i\phi_i}-e^{i\phi_j}|^{\beta\over 2}\;
, \label{eq:13}
\end{equation}
where $C$, the normalization constant, is the same as that of
Sutherland ground state wave function. Note that since $\alpha^2$
in Eq.~(\ref{eq:8}) has to be positive, the equivalence holds
only for $\beta\ge 2$. If single-valuedness of $\Psi_0$ is
demanded under the transformation $\phi_i\rightarrow(\phi_i+2\pi)$,
then further restrictions apply on the allowed values of $\alpha$.
It is also interesting  to note that $E_0$ in Eq.~(\ref{eq:12})
has a nonlinear dependence on the parameter $\alpha$. As in the
Sutherland model, the thermodynamics (and the virial coefficients)
 of the anyons on a ring can be easily worked out in the limit
$L\rightarrow \infty$.

This research is supported by  grants from the Natural Sciences and
Engineering
Research Council of Canada.
The authors would like to thank Jim Waddington for discussions
and support. Thanks are also due to Avinash Khare and M.V.N. Murthy.

\end{document}